\documentclass[twocolumn,A4]{article}
\usepackage[dvips]{graphics}
\begin{document}
\onecolumn
\begin{center}
{\bf{\Large Anomalous quantum transport in a thin film}}\\
~\\
Santanu K. Maiti$^{\dag,\ddag,*}$ \\
~\\
{\em $^{\dag}$Theoretical Condensed Matter Physics Division,
Saha Institute of Nuclear Physics, \\
1/AF, Bidhannagar, Kolkata-700 064, India \\
$^{\ddag}$Department of Physics, Narasinha Dutt College,
129, Belilious Road, Howrah-711 101, India} \\
~\\
{\bf Abstract}
\end{center}
We present a numerical study of electron transport in a thin film
of varying disorder strength with the distance from its surface. 
A simple tight-binding model is used to describe the system, in
which the film is attached to two metallic electrodes and the 
coupling of this film to the electrodes is illustrated by the
Newns-Anderson chemisorption theory. Quite interestingly we observe
that, in the smoothly varying disordered film current amplitude
increases with the increase of disorder strength in the strong
disorder regime, while it decreases in the weak disorder regime.
This behavior is completely opposite from a conventional disordered
film, where current amplitude always decreases with the increase
of disorder strength. 
\vskip 1cm
\begin{flushleft}
\end{flushleft}
\vskip 4.5in
\noindent
{\bf ~$^*$Corresponding Author}: Santanu K. Maiti

Electronic mail: santanu.maiti@saha.ac.in
\newpage
\twocolumn

\section{Introduction}

Improvements of nanoscience and technology have stimulated us to investigate
electrical conduction on mesoscopic/nanoscopic scale in a very tunable
environment. The transport properties of quantum systems attached to
electrodes have been studied extensively over the last few decades both
theoretically as well as experimentally due to their possible technological
applications. In $1974$, Aviram and Ratner$^1$ first developed a theoretical
formulation for the description of electron conduction in a molecular
electronic device. Later many experiments$^{2-6}$ have been performed in
several bridge systems to justify the basic mechanisms underlying the
electron transport. From the numerous studies of electron transport,
exist in the literature, we can able to understand different features,
but yet the complete knowledge of the conduction mechanism in this
scale is not well established even today. Many significant factors are
there which control the electron transport in a molecular bridge, and
all these effects have to be taken into account properly to reveal the
transport. For our illustrative purpose, here we mention some of them
as follows. In order to reveal the dependence of the molecular structure
on the electron transport, Ernzerhof {\em et al.}$^7$ have performed 
few model calculations and predicted some interesting results. The
molecular coupling$^8$ to the side attached electrodes is another
important parameter that controls the electron transport in a 
significant way. The most significant issue is probably the effects
of quantum interferences of electron waves in different pathways,
and several studies$^{8-16}$ are available in the literature describing 
these effects. In addition to these, dynamical fluctuations provide an
active role in the determination of molecular transport which can be
manifested through the measurement of {\em shot noise}, a direct
consequence of the quantization of charge. This can be used to obtain 
information on a system which is not directly available through 
conductance measurements, and is generally more sensitive to the effects 
of electron-electron correlations than the average conductance.$^{17,18}$

In this article, we concentrate ourselves on a different aspect, related 
to the effect of disorder, of quantum transport than the above 
mentioned issues. The characteristic properties of electron transport
in conventional disordered systems are well established in the 
literature. But there are some special type of nano-scale materials,
where the charge carriers are scattered mainly from their surface
regions$^{19-23}$ and not from the inner core regions. These systems
exhibit several peculiar features in electron transport. For example,
in a very recent work Yang {\em et al.}$^{21}$ have shown a 
localization to quasi-delocalization transition in edge disordered
graphene nanoribbons by varying the strength of edge disorder.
Quite similar in behavior has also been observed in a shell-doped
nanowire,$^{22}$ where the electron dynamics undergoes a localization
to quasi-delocalization transition beyond some critical doping.
In the shell-doped nanowire, the dopant atoms are spatially
confined within a few atomic layers in the shell region. This is
completely opposite to that of a traditional disordered nanowire,
where the dopant atoms are distributed uniformly throughout the system.
From the numerous studies of electron transport in many such systems,
it can be emphasized that the surface reconstructions,$^{24}$
surface scattering$^{25}$ and surface states$^{26}$ may be quite
significant to exhibit several diverse transport properties.
Motivated with these investigations, here we consider a particular
kind of thin film in which disorder strength varies smoothly
from layer to layer with the distance from its surface. In this
system, our numerical study predicts a strange behavior of electron
transport where current amplitude increases with the increase of 
disorder strength in the strong disorder limit, while it decreases 
in the limit of weak disorder. On the other hand, for a traditional
disordered thin film current amplitude always decreases with the 
increase of disorder strength. An analytic approach based on the
tight-binding model is used to incorporate the electron transport
in the film, and we adopt the Newns-Anderson chemisorption 
model$^{27-29}$ to describe the side attached electrodes and the
interaction of these electrodes with the film.

Our organization of this article is as follows. In Section $2$, we 
describe the model and the theoretical formulation for our calculations. 
Section $3$ discusses the significant results, and at the end, we 
summarize our results in Section $4$.

\section{Model and the theoretical description}

\begin{figure}[ht]
{\centering \resizebox*{7cm}{4cm}{\includegraphics{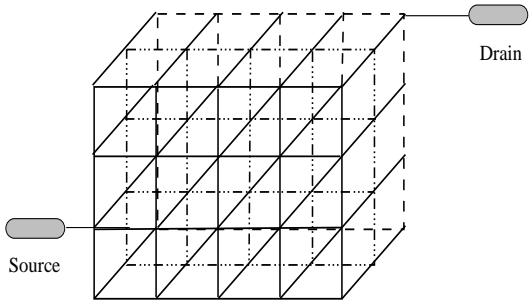}}\par}
\caption{Schematic view of a smoothly varying disordered thin film attached 
to two metallic electrodes (source and drain). The top most front layer 
(solid line) is the highest disordered layer and the disorder strength 
decreases smoothly to-wards the bottom layer keeping the lowest bottom 
layer (dashed line) as disorder free. Two electrodes are attached at the 
two extreme corners of the bottom layer.}
\label{quantumfilm}
\end{figure}

The system of our concern is depicted in Fig.~\ref{quantumfilm}, where
a thin film is attached to two metallic electrodes, viz, source and
drain. In this film, disorder strength varies smoothly from the
top most disordered layer (solid line) to-wards the bottom layer,
keeping the lowest bottom layer (dashed line) as disorder free.
The electrodes are symmetrically attached at the two extreme corners 
of the bottom layer. Using the Green's function formalism$^{30}$
and single channel Landauer conductance formula,$^{30}$ we 
calculate the transmission probability ($T$), conductance ($g$)
and current ($I$) through the film.  

At low temperature and bias voltage, the conductance $g$ of the film is 
obtained from the Landauer conductance formula,$^{30}$
\begin{equation}
g=\frac{2e^2}{h} T
\label{equ1}
\end{equation}
where the transmission probability $T$ can be written in terms of the
retarded and advanced Green's functions ($G_F^r$ and $G_F^a$) of the 
film as$^{30}$
\begin{equation}
T=Tr\left[\Gamma_S G_F^r \Gamma_D G_F^a\right]
\label{equ2}
\end{equation}
The parameters $\Gamma_S$ and $\Gamma_D$ describe the couplings of the film 
with the source and drain respectively. 

The Green's function of the film becomes,
\begin{equation}
G_{F}=\left(E-H_{F}-\Sigma_S-\Sigma_D\right)^{-1}
\label{equ3}
\end{equation}
where $E$ is the energy of the injecting electron and $H_{F}$ corresponds
to the Hamiltonian of the film. The parameters $\Sigma_S$ and $\Sigma_D$
denote the self-energies due to the coupling of the film with the source
and drain respectively, and are described by the use of Newns-Anderson
chemisorption theory.$^{27-29}$ In the tight-binding framework, the
Hamiltonian can be expressed within the non-interacting picture as,
\begin{equation}
H_{F}=\sum_i \epsilon_i c_i^{\dagger} c_i + \sum_{<ij>} t \left(c_i^{\dagger}
c_j + c_j^{\dagger} c_i\right)
\label{equ4}
\end{equation}
Here $\epsilon_i$ gives the on-site energy of an electron at site $i$ and
$t$ represents the hopping strength between two nearest-neighbor sites
both for the longitudinal and transverse directions of the film.
Now to achieve our concerned system, we choose the site energies 
($\epsilon_i$'s) randomly from a ``Box" distribution function such that 
the top most front layer becomes the highest disordered layer with 
strength $W$ and the strength of disorder decreases smoothly to-wards 
the bottom layer as a function of $W/(N_l-m)$, where $N_l$ gives the 
total number of layers and $m$ represents the total number of ordered 
layers from the bottom side of the film. 
While, in the conventional disordered thin film all the layers are 
subjected to the same disorder strength $W$. In our present model, the 
electrodes are described by the standard tight-binding Hamiltonian, 
similar to that as prescribed in Eq.(\ref{equ4}), and parametrized by 
constant on-site potential $\epsilon_0$ and nearest-neighbor hopping 
integral $v$.

Assuming the entire voltage is dropped across the film-electrode
interfaces,$^{31}$ the current passing through the film, which is 
regarded as a single electron scattering process between the 
reservoirs, can be expressed as,$^{30}$
\begin{equation}
I(V)=\frac{e}{\pi \hbar}\int_{-\infty}^{\infty} \left(f_S-f_D\right) 
T(E) dE
\label{equ5}
\end{equation}
where $f_{S(D)}=f\left(E-\mu_{S(D)}\right)$ gives the Fermi distribution
function with the electrochemical potential $\mu_{S(D)}=E_F\pm eV/2$.
In this article, we focus our study on the determination of the 
typical current amplitude which is obtained from the relation,
\begin{equation}
I_{typ}=\sqrt{<I^2>_{W,V}}
\label{equ6}
\end{equation}
where $W$ and $V$ correspond to the impurity strength and the applied 
bias voltage respectively.

Throughout this presentation, all the results are computed at absolute 
zero temperature, but they should valid even for finite temperature 
($\sim 300$ K) as the broadening of the energy levels of the film due 
to its coupling with the electrodes will be much larger than that of
the thermal broadening.$^{30}$ For simplicity, we take the unit $c=e=h=1$ 
in our present calculations.

\section{Results and discussion}

All the numerical calculations we do here are performed for some
particular values of the different parameters, but all the basic
features in which we are interested in this particular study remain
also invariant for the other parametric values. The values of the
different parameters are as follows. The coupling strengths of the
film to the electrodes are taken as $\tau_S=\tau_D=1.5$, the 
nearest-neighbor hopping integral in the film is fixed to $t=1$. 
The on-site potential and the hopping integral in the electrodes are 
set as $\epsilon_0=0$ and $v=2$ respectively. In addition to these,
here we also introduce three other parameters $N_x$, $N_y$ and $N_z$ to
specify the system size of the thin film, where they correspond to the
total number of lattice sites along the $x$, $y$ and $z$ directions
of the film respectively. In our numerical calculations, the typical
current amplitude ($I_{typ}$) is determined by taking the average
over the disordered configurations and bias voltages (see Eq.(\ref{equ6})).
Since in this particular model the site energies are chosen randomly,
we compute $I_{typ}$ by taking the average over a large number ($60$)
of disordered configurations in each case to get much accurate result. 
On the other hand, for the averaging over the bias voltage $V$, we set
the range of it from $-10$ to $10$. In this presentation, we focus
only on the systems with small sizes since all the qualitative behaviors 
remain also invariant even for the large systems.

\begin{figure}[ht]
{\centering \resizebox*{7cm}{5cm}{\includegraphics{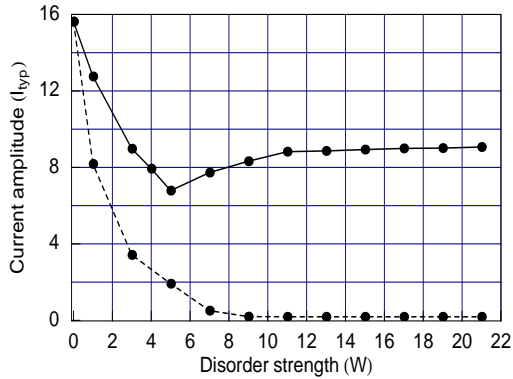}}\par}
\caption{$I_{typ}$ vs. $W$ for some thin films with $N_x=10$, $N_y=8$
and $N_z=5$. Here we set $m=1$. The solid and dotted curves correspond 
to the smoothly varying and complete disordered films respectively.}
\label{disorder2}
\end{figure}

Figure~\ref{disorder2} illustrates the variation of the typical current
amplitude ($I_{typ}$) as a function of the disorder ($W$) for some thin
films with $N_x=10$, $N_y=8$ and $N_z=5$. Here we set $m=1$, i.e., only 
the lowest bottom layer is free from any disorder for these films. 
The solid and dotted curves correspond to the results of the smoothly 
varying and complete disordered
thin films respectively. A remarkably different behavior is observed 
for the smoothly varying disordered film compared to the film with
complete disorder. In the later system, it is observed that $I_{typ}$
decreases rapidly with $W$ and eventually it drops to zero for the
higher value of $W$. This reduction of the current is due to the
fact that the eigenstates become more localized$^{32}$ with the increase 
of disorder, and it is well established from the theory of Anderson
localization.$^{33}$ The appreciable change in the variation of the 
typical current amplitude takes place only for the unconventional
disordered film. In this case, the current amplitude decreases 
initially with $W$ and after reaching to a minimum at $W=W_c$ (say),
it again increases. Thus the anomalous behavior is observed beyond the 
critical disorder strength $W_c$, and we are interested particularly
in this regime where $W>W_c$. In order to illustrate this peculiar
behavior, we consider the smoothly varying disordered film as a coupled
system combining two sub-systems. The coupling exists between the 
lowest bottom ordered layer and the other disordered layers. Thus
the system can be treated, in other way, as a coupled order-disorder
separated thin film. For this coupled system we can write the
Schr\"{o}dinger equations as: $(H_0-H_1)\psi_0=E\psi_0$ and
$(H_d-H_2)\psi_d=E\psi_d$. Here $H_0$ and $H_d$ represent the
sub-Hamiltonians of the ordered and disordered regions of the film
respectively, and $\psi_0$ and $\psi_d$ are the corresponding
eigenfunctions. The terms $H_1$ and $H_2$ in the above two
expressions are the most significant and they can be expressed as:
$H_1=H_{od}(H_d-E)^{-1}H_{do}$ and $H_2=H_{do}(H_o-E)^{-1}H_{od}$.
$H_{od}$ and $H_{do}$ correspond to the coupling between the ordered
region and the disordered region. From these mathematical expressions,
the anomalous behavior of the electron transport in the film
can be described clearly. In the absence of any interaction between 
the ordered and disordered regions, we can assume the full system as
a simple combination of two independent sub-systems. Therefore, we get 
all the extended states in the ordered region, while the localized 
states are obtained in the disordered region. In this situation, the 
motion of an electron in any one region is not affected by the other. 
But for the coupled system, the motion of the electron is no more
independent, and we have to take the combined effects coming from both 
the two regions. With the increase of disorder, the scattering 
effect becomes dominated more, and thus the reduction of the current 
is expected. This scattering is due to the existence of the localized 
eigenstates in the disordered regions. Therefore, in the case of
strong coupling between the two sub-systems, the motion of the electron
in the ordered region is significantly influenced by the disordered
regions. Now the degree of this coupling between the two sub-systems
solely depends on the two parameters $H_1$ and $H_2$, those are expressed 
earlier. In the limit of weak disorder, the scattering effect from both
the two regions is quite significant since then the terms $H_1$ and $H_2$ 
have reasonably high values. With the increase of disorder,
$H_1$ decreases gradually and for a very large value of $W$ it becomes
very small. Hence the term $(H_0-H_1)$ effectively goes to $H_0$ in the 
limit $W \rightarrow 0$, which indicates that the ordered region becomes 
decoupled from the disordered one. Therefore, in the higher disorder
\begin{figure}[ht]
{\centering \resizebox*{7cm}{5cm}{\includegraphics{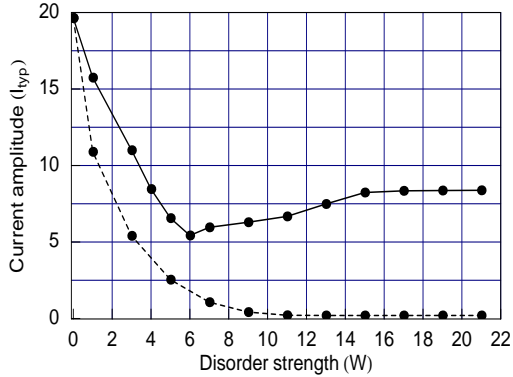}}\par}
\caption{$I_{typ}$ vs. $W$ for some thin films with $N_x=12$, $N_y=10$
and $N_z=6$. Here we set $m=2$. The solid and dotted curves correspond to 
the identical meaning as in Fig.~\ref{disorder2}.}
\label{disorder3}
\end{figure}
regime the scattering effect becomes less significant from the ordered
region, and it decreases
with $W$. For the low regime of $W$, the eigenstates of both the two
effective Hamiltonians, $(H_0-H_1)$ and $(H_d-H_2)$, are localized.
With the increase of $W$, $H_1$ gradually decreases, resulting in 
much weaker localization in the states of $(H_0-H_1)$, while the states
of $(H_d-H_2)$ become more localized. At a critical value of $W=W_c$
(say) ($\simeq$ band width of $H_0$), we get a separation between 
the much weaker localized states and the strongly localized states.
Beyond this value, the weaker localized states become more extended
and the strongly localized states become more localized with the
increase of $W$. In this situation, the current is obtained mainly 
from these nearly extended states which provide the larger current
with $W$ in the higher disorder regime.

To illustrate the size dependence of the film on the electron transport,
in Fig.~\ref{disorder3} we plot the variation of the typical current
amplitude for some thin films with $N_x=12$, $N_y=10$ and $N_z=6$. For 
these films we take $m=2$, i.e., two layers from the bottom side are
free from any disorder. The solid and dotted curves correspond to the
identical meaning as in Fig.~\ref{disorder2}. For both the unconventional 
and traditional disordered films, we get almost the similar behavior of
the current as presented in Fig.~\ref{disorder2}. This study shows that 
the typical current amplitude strongly depends on the finite size of the 
thin film.

\section{Concluding Remarks}

To summarize, we have done a numerical study to show the anomalous
behavior of electron transport in a unconventional disordered thin 
film where the disorder strength varies smoothly from its surface.
A simple tight-binding model has been used to describe the system,
where the coupling of the film to the electrodes has been described
by the Newns-Anderson chemisorption theory. We have calculated the
typical current amplitudes by using the Green's function formalism,
and our results have shown a remarkably different behavior for the
unconventional disordered film compared to the traditional disordered
film. In the smoothly varying disordered film, the typical current
amplitude decreases with $W$ in the weak disorder regime ($W<W_c$),
while it increases in the strong disorder regime ($W>W_c$). On the
other hand for the conventional disordered film, the current amplitude
always decreases with disorder. In this present investigations, we
have also studied the finite size effects of the film and our numerical
results have shown that the typical current amplitude strongly
depends on the size of the film. Quite similar feature of anomalous 
quantum transport can also be observed in some other lower dimensional 
systems like, edge disordered graphene sheets of single-atom-thick, 
finite width rings with surface disorder, nanowires, etc. Our study has 
revealed that the carrier transport in an order-disorder separated 
mesoscopic device may be tailored to desired properties through doping 
for different applications.

\end{document}